\documentclass[twocolumn,amsmath,amssymb,prb,longbibliography]{revtex4-1}
\usepackage{multirow}
\usepackage{array}
\usepackage{amsmath}
\usepackage{graphicx}
\usepackage{dcolumn}
\usepackage{bm}
\usepackage{verbatim}
\usepackage{comment}
\DeclareMathAlphabet \mathbfcal{OMS}{cmsy}{b}{n}
\begin{document}



\title{Ultrafast optical currents in gapped graphene}

\author{S. Azar Oliaei Motlagh}
\author{Fatemeh Nematollahi}
\author{Aranyo Mitra}
\author{Jawad Zafar}
\author{Vadym Apalkov}
\author{Mark I. Stockman}
\affiliation{Center for Nano-Optics (CeNO) and
Department of Physics and Astronomy, Georgia State
University, Atlanta, Georgia 30303, USA
}

\date{\today}
\begin{abstract}
We study theoretically the interaction of ultrashort optical pulses with gapped graphene. Such strong pulse results in finite conduction band population and 
corresponding electric current both 
during and after the pulse. Since gapped graphene has broken inversion symmetry, it has an axial symmetry about the $y$-axis but not about the $x$-axis. 
We show that, in this case, 
if the linear pulse is polarized along the $x$-axis, the rectified electric current is generated in the $y$ direction. At the same time, the conduction band 
population distribution in the reciprocal space is symmetric about the $x$-axis. Thus, the rectified current in gapped graphene has inter-band 
origin, while the intra-band contribution to the rectified current is zero.


\end{abstract}
\maketitle
\section{Introduction}

The availability of ultrashort laser pulses with the duration of a few femtoseconds provides effective tools to manipulate and study the electron dynamics in solids at ultrafast time scale with high temporal resolution\cite{Schiffrin_at_al_Nature_2012_Current_in_Dielectric, Apalkov_Stockman_PRB_2012_Strong_Field_Reflection, Higuchi_Hommelhoff_et_al_Nature_2017_Currents_in_Graphene, Gruber_et_al_ncomms13948_2016_Ultrafast_pulses_graphene, Stockman_et_al_PhysRevB.95_2017_Crystalline_TI,Stockman_et_al_PhysRevB.98_2018_3D_TI,Stockman_et_al_PhysRevB.99_2019_Weyl, Hommelhoff_et_al_PhysRevLett.121_2018_Coherent, Ghimire_et_al_Nature_Communications_2017_HHG, Reis_et_al_Nat_Phys_2017_HHG_from_2D_Crystals, Simon_et_al_PRB_2000_Strong_Field_Fs_Ionization_of_Dielectrics, Hommelhoff_et_al_1903.07558_2019_laser_pulses_graphene, sun_et_al_nnano.2011.243_2012_Ultrafast_pulses_graphene, Mashiko_et_al_Nature_Communications_2018_ultrafast_pulse_solid, Shin_et_al_IOP_Publishing_2018_ultrafast_pulse_solid,Hommelhoff_et_al_PhysRevLett.121_2018_Coherent, 
Gruber_et_al_ncomms13948_2016_Ultrafast_pulses_graphene, Higuchi_et_al_Nature_2017, Leitenstorfer_et_al_PhysRevB.92_2015_Ultrafast_Pseudospin_Dynamics_in_Graphene, Stockman_et_al_PhysRevB.96_2017_Berry_Phase, Stockman_et_al_PhysRevB.98_2018_Rapid_Communication_Topological_Resonances, Sun_et_al_Chinese_Physics_B_2017_Ultrafast_pulses_TMDC, Zhang_et_al_OSA_2018_ultrafast_pulse_TMDC}. Among solids two dimensional (2D) crystalline materials exhibits unique properties due to the confinement of electron dynamics to a plane \cite{Butler_et_al_Acs_Nano_2013_2D_Beyond_Graphene}. 
 Graphene, a layer of carbon atoms with the thickness of one atom, is well known 2D material with fascinating properties. Graphene has a honeycomb crystal structure made of two sublattices, A and B - see Fig.\ \ref{fig:Energy}(a)\cite{Geim_et_al_Nat_Mater_2007_The_rise_of_graphene,Electronic_properties_graphene_RMP_2009}. Having two Dirac points, $K^\prime$ and $K$ at the edges of the Brillouin zone -see Fig.\ \ref{fig:Energy}(b), makes graphene a suitable platform to study the dynamics of massless Dirac fermions \cite{Butler_et_al_Acs_Nano_2013_2D_Beyond_Graphene,Novoselov_Geim_et_al_nature04233_2D_Electrons_in_Graphene,Geim_et_al_Nat_Mater_2007_The_rise_of_graphene,Electronic_properties_graphene_RMP_2009}. In graphene, both time reversal and inversion symmetries are conserved. 
 However, there is a broad class of semiconductors with honeycomb crystal structure where two sublattices are made of two different atoms, and the inversion symmetry is broken, which results in a finite bandgap at the $K$ and $K^\prime$ points \cite{Kormanyos_et_al_2d_Materials_2015_k.p_theory_for_two_dimensional,Jiang_Frontiers_of_Physics_2015_Graphene_versus_MoS2}. One of such materials is a monolayer of transition metal dichalcogenides (TMDCs) that has a direct bandgap with nonzero Berry curvature around the $K^\prime$ and $K$ valleys. Gapped graphene, which has broken inversion 
 symmetry, has topological properties similar to TMDC monolayer. Namely, the Berry curvature in gapped graphene is extended over the finite region near the $K$ and $K^\prime $ points. Such broadening of the Berry curvature, which can be tuned by the bandgap, results in nontrivial topological properties of gapped graphene   \cite{Kormanyos_et_al_2d_Materials_2015_k.p_theory_for_two_dimensional,Ye_et_al_Nature_Nanotechnology_2016_Electrical_generation_and_control, Sun_et_al_Chinese_Physics_B_2017_Ultrafast_pulses_TMDC,Jariwala_et_al_Asc_Nano_2014_Transition_Metal}. One of such properties is recently predicted topological resonance, which produces finite valley polarization in transition metal dichalcogenides and gapped graphene\cite{Stockman_et_al_PhysRevB.98_2018_Rapid_Communication_Topological_Resonances}.
 
In this article, we study the ultrafast nonlinear electron dynamics in gapped graphene. The dynamics is induced by a single cycle ultrafast linearly polarized pulse. 
Although the linear pulse does not produce any residual valley polarization, it results in electric current, the magnitude and the direction of which can be 
controlled by the bandgap. 
Gapped graphene, considered in the present article, is a model of direct bandgap semiconductors with honeycomb lattice structures. Opening of the bandgap in graphene can be achieved by several methods, for example, by placing graphene on Boron Nitride (BN) or silicon carbide (SiC) substrate\cite{Conrad_et_al_PhysRevLett.115_2015_Gapped_Graphene_on_SiC,Ajayan_et_al_Review-JNN_2011_Band_Opening_in_Graphene}. 


\section{MODEL AND MAIN EQUATIONS}
\label{Model_and_Equations}

\begin{figure}
\begin{center}\includegraphics[width=0.47\textwidth]{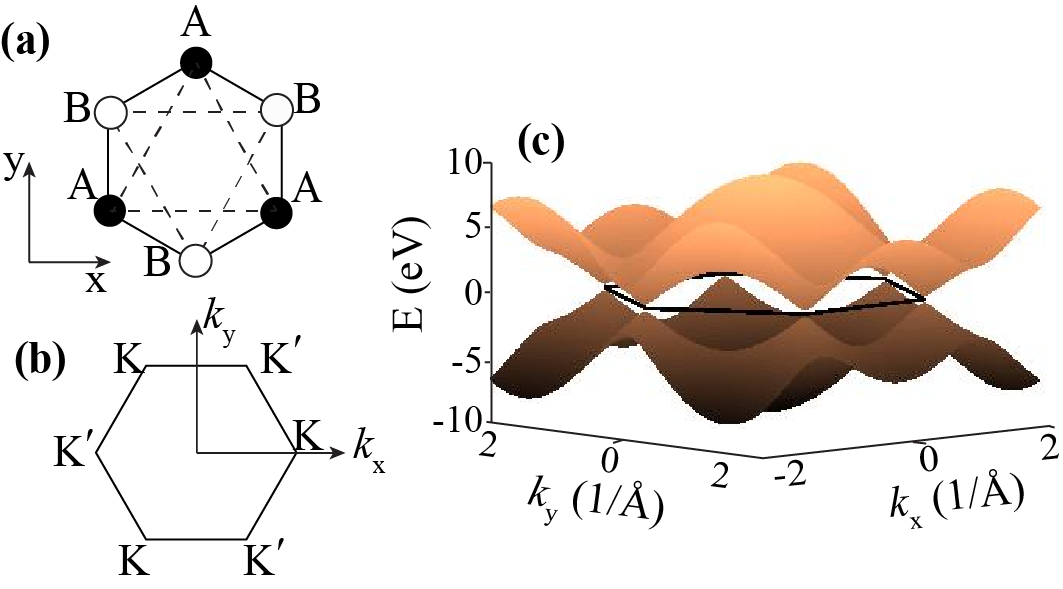}\end{center}
  \caption{(Color online) (a) The honeycomb lattice structure of graphene is made of two triangular sublattices A (black circle) and B (white circle). (b) The first Brillouin zone of the honeycomb lattice has two valleys $K$ and $K^\prime$ located in its boundaries. (c) The energy dispersion is shown for gapped graphene with the bandgap of 0.5 eV in the extended zone. The solid black lines show the boundaries of the first Brillouin zone. }
  \label{fig:Energy}
\end{figure}

In the presence of an applied ultrafast optical pulse, $\mathbf F(t)$, with the duration of less than 5 fs, the electron dynamics is coherent. This assumption is valid since the electron scatering time in  2D materials is longer than 10 fs \cite{Hwang_Das_Sarma_PRB_2008_Graphene_Relaxation_Time, Breusing_et_al_Ultrafast-nonequilibrium-carrier-dynamics_PRB_2011, theory_absorption_ultrafast_kinetics_graphene_PRB_2011, Ultrafast_collinear_scattering_graphene_nat_comm_2013, Gierz_Snapshots-non-equilibrium-Dirac_Nat-Material_2013, Nonequilibrium_dynamics_photoexcited_electrons_graphene_PRB_2013}. To find the coherent electron dynamics in gapped graphene we solve time-dependent Schr\"odinger equation (TDSE)  
\begin{equation}
i\hbar \frac{{d\Psi }}{{dt}} = { H(t)} \Psi,  
\label{Sch}
\end{equation}
with the Hamiltonian
\begin{equation}
{ H}(t) = { H}_0 - e{\bf{F}}(t){\bf{r}},
\label{Ht}
\end{equation}  
where $e$ is an electron charge, and $H_0$ is the nearest neighbor tight binding Hamiltonian of gapped graphene \cite{Kjeld_et_al_PhysRevB.79.113406_2009_Gapped_Graphene_Optical_Response},
\begin{eqnarray}
H_0=\left( {\begin{array}{cc}
   \frac{\Delta}{2} & \gamma f(\mathbf k) \\
   \gamma f^\ast(\mathbf k) & -\frac{\Delta}{2} \\
  \end{array} } \right) .
\label{H0}
\end{eqnarray}
Here $\Delta$ is the bandgap, 
$\gamma= -3.03$ eV is the hopping integral, and
\begin{equation}    
f(\mathbf k)=\exp\Big(i\frac{ak_y}{\sqrt{3}}\Big )+2\exp\Big(-i\frac{ak_y}{2\sqrt{3}}\Big )\cos{\Big(\frac{ak_x}{2}\Big )},
\end{equation}
where $a=2.46~\mathrm{\AA}$ is a lattice constant. The eigenenergies of the tight-binding Hamiltonian, $H_0$, can be found as follows
\begin{eqnarray}
E_{c}(\mathbf k)&=&+\sqrt{\gamma ^2\left |{f(\mathbf k)}\right |^2+\frac{\Delta ^2}{4} }~~,
\label{Energyc} \\
E_{v}(\mathbf k)&=& -\sqrt{\gamma ^2\left |{f(\mathbf k)}\right |^2+\frac{\Delta ^2}{4}}~~,
\label{Energy}
\end{eqnarray}
where $c$ and $v$ stand for the conduction band (CB) and the valence band (VB), respectively. Figure \ref{fig:Energy}(c) shows the calculated energy dispersion from Eqs.\ (\ref{Energyc}) and (\ref{Energy}) for the bandgap of $\Delta=0.5~ \mathrm{eV}$.

The coherent electron dynamics in solids has two major components: intraband and interband dynamics. The intraband dynamics is governed by the Bloch acceleration theorem 
\begin{equation}
\hbar \frac{{d{\bf{k}}}}{{dt}} = e{\bf{F}}(t).
\label{acceleration}
\end{equation}
The solution of this equation has the following form
\begin{equation}
{{\bf{k}}}({\bf{q}},t) = {\bf{q}} + \frac{e}{\hbar }\int_{ - \infty }^t {{\bf{F}}({t^\prime})d{t^\prime}}, 
\label{kvst}
\end{equation}
where ${\bf q}$ is the initial crystal wavevector of an electron in the first Brillouin zone. 

The corresponding wave functions, which are the solutions of Schr\"odinger equation (\ref{Sch}) within a single band $\alpha$, i.e., without interband coupling, are the Houston functions \cite{Houston_PR_1940_Electron_Acceleration_in_Lattice},
\begin{equation}
 \Phi^\mathrm{(H)}_{\alpha {\bf q}}({\bf r},t)=\Psi^{(\alpha)}_{\bf{k}(\bf q,t)} ({\bf r})\exp\left(i\phi^{(\mathrm d)}_{\mathrm{\alpha}}({\bf q},t)+i\phi^{(\mathrm B)}_{\mathrm{\alpha}}({\bf q},t)\right),
\end{equation}
where $\alpha=v,c$ stand for the VB and CB, respectively,  $ \mathrm{\Psi^{(\alpha)}_{{\mathbf k}}} $ are Bloch-band eigenstates in the absence of the external field, $E_\alpha(\mathbf k)$ are the eigenenergies, and the dynamic phase, $\phi^\mathrm{(D)}_{\mathrm \alpha}$, and geometric phase, $\phi^\mathrm{(B)}_{\mathrm \alpha}$, are defined as 
\begin{eqnarray}
\phi^\mathrm{(D)}_{\alpha}(\mathbf q,t)= \frac{-1}{\hbar} \int_{-\infty}^t dt^\prime \left(E_\mathrm \alpha[\mathbf k (\mathbf q,t^\prime)]\right),
 \label{phi}
 \\ 
 \phi^\mathrm{(B)}_{\mathrm \alpha}(\mathbf q,t)= \frac{e}{\hbar} \int_{-\infty}^t dt^\prime \mathbf F \left(\mathbfcal{A}^{\mathrm{\alpha \alpha}}[\mathbf k (\mathbf q,t^\prime)]\right).
 \label{phi}
\end{eqnarray}
Here  $\mathbfcal{A}^{\alpha\alpha}=\left\langle \Psi^{(\alpha)}_\mathbf q  |   i\frac{\partial}{\partial\mathbf q}|\Psi^{(\alpha)}_\mathbf q   \right\rangle $ is the intraband Berry connection. The expressions for the intraband Berry connections, $\mathbfcal{A}^{\alpha\alpha}=(\mathcal{A}^{\alpha\alpha}_x,\mathcal{A}^{\alpha\alpha}_y)$, 
can be found from the tight-binding Hamiltonian as follows
\begin{eqnarray}
\mathcal{A}_{x}^{\alpha\alpha}(\mathbf k)&=&\frac{-a\gamma ^2}{\gamma ^2 |f(\mathbf k)|^2+(\Delta/2-E_\alpha)^2}
 \sin \frac{3ak_y}{2\sqrt{3}}\sin{\frac{ak_x}{2}},
\nonumber \\
 \label{Ax_alpha}
 \\
\mathcal{A}_{y}^{\alpha\alpha}(\mathbf k)&=&\frac{a\gamma ^2}{\sqrt{3}\left(\gamma ^2 |f(\mathbf k)|^2+(\Delta/2-E_\alpha)^2\right)}\nonumber\\
 &&\times \left(\cos{ ak_x}-\cos{\frac{\sqrt{3}ak_y}{2}}\cos{\frac{ak_x}{2}}\right).
 \label{Ay_alpha}
\end{eqnarray}

The interband electron dynamics is described by TDSE (\ref{Sch}). The solution of TDSE can be expanded in the basis of Houston functions $\Phi^{(H)}_{\alpha {\bf q}}({\bf r},t)$ \cite{Houston_Phys_Rev_1940_Acceleration_of_Electrons_in_a_Crystal_Lattice},
\begin{equation}
\Psi_{\bf q} ({\bf r},t)=\sum_{\alpha=c,v}\beta_{\alpha{\bf q}}(t) \Phi^{(H)}_{\alpha {\bf q}}({\bf r},t),
\end{equation}
where 
$\beta_{\alpha{\bf q}}(t)$ are expansion coefficients, which satisfies the following system of coupled differential equations
\begin{equation}
i\hbar\frac{\partial B_\mathbf q(t)}{\partial t}= H^\prime(\mathbf q,t){B_\mathbf q}(t)~,
\label{Schrodinger}
\end{equation}
where the wave function (vector of state) $B_q(t)$ and Hamiltonian $ H^\prime(\mathbf q,t)$ are defined as 
\begin{eqnarray}
B_\mathbf q(t)&=&\begin{bmatrix}\beta_{c\mathbf q}(t)\\ \beta_{v\mathbf q}(t)\\ \end{bmatrix}~,\\ 
H^\prime(\mathbf q,t)&=&-e\mathbf F(t)\mathbfcal{\hat A}(\mathbf q,t)~,\\
\mathbfcal{\hat A}(\mathbf q,t)&=&\begin{bmatrix}0&\mathbfcal D^{cv}(\mathbf q,t)\\
\mathbfcal D^{vc}(\mathbf q,t)&0\\
\end{bmatrix}~.
\end{eqnarray}
where
\begin{eqnarray}
\mathbfcal D^{cv}(\mathbf q,t)&=&
\mathbfcal A^{cv}[\mathbf k (\mathbf q,t)]\nonumber\\
&\times & \exp\left(i\phi^\mathrm{(D)}_{cv}(\mathbf q,t)+i\phi^\mathrm{(B)}_{cv}(\mathbf q,t)\right),
 \label{Q}
\\
\phi^\mathrm{(D)}_{cv}(\mathbf q,t)&=&\phi^\mathrm{(D)}_{v}(\mathbf q,t)-\phi^\mathrm{(D)}_{c}(\mathbf q,t)
 \label{phiD}
 \\
\phi^\mathrm{(B)}_{cv}(\mathbf q,t)&=&\phi^\mathrm{(B)}_{v}(\mathbf q,t)-\phi^\mathrm{(B)}_{c}(\mathbf q,t)
 \label{phiB}
 \\ 
{\mathbfcal{A}}^{cv}({\mathbf q})&=&
\left\langle \Psi^{(c)}_\mathbf q  |   i\frac{\partial}{\partial\mathbf q}|\Psi^{(v)}_\mathbf q   \right\rangle .
\label{D}
\end{eqnarray} 
Here  ${\mathbfcal A}^{cv}(\mathbf q)$ is  a matrix element of the non-Abelian Berry connection \cite{Wiczek_Zee_PhysRevLett.52_1984_Nonabelian_Berry_Phase, Xiao_Niu_RevModPhys.82_2010_Berry_Phase_in_Electronic_Properties, Yang_Liu_PhysRevB.90_2014_Non-Abelian_Berry_Curvature_and_Nonlinear_Optics},
which has the following expression
\begin{eqnarray}
\mathcal{A}_{x}^{cv}(\mathbf k)&=&\mathcal N\Bigg(\frac{-a}{2|f(\mathbf k)|^2}\Bigg)\Bigg( \sin\frac{ak_x}{2}\sin\frac{a\sqrt{3}k_y}{2}
\nonumber\\
&&+i \frac{\Delta}{2E_c}\Bigg(\cos \frac{a\sqrt{3}k_y}{2}\sin \frac{ak_x}{2}+\sin{ak_x}\Bigg)\Bigg)
\nonumber \\
 \label{Ax}
\\
\mathcal{A}_{y}^{cv}(\mathbf k)&=&\mathcal N\Bigg(\frac{a}{2\sqrt{3}|f(\mathbf k)|^2}\Bigg)\Bigg( -1-\cos\frac{a\sqrt{3}k_y}{2}\cos\frac{ak_x}{2}
\nonumber\\
&&+2\cos ^2 \frac{ak_x}{2}-i \frac{3\Delta}2{E_c}\sin \frac{a\sqrt{3}k_y}{2}\cos \frac{ak_x}{2}\Bigg),
\nonumber \\
\label{Ay}
\end{eqnarray}
where
\begin{equation}
\mathcal N=\frac{\left|\gamma f(\mathbf k)\right|}{\sqrt{\frac{\Delta ^2}{4}+\left|\gamma f(\mathbf k)\right|^2}}~.\nonumber
\end{equation}

The ultrafast field drives electric current, ${\mathbf J}(t) = \left\{J_x(t),J_y(t)\right\}$. The current has both interband and intraband contributions, $\mathbf J(t)=\mathbf J^\text{(intra)}(t)+\mathbf J^\text{(inter)}(t)$. The intraband current is proportional to the group velocity and has the following form
\begin{equation}
\mathbf J^\text{(intra)}(t)  =\frac{eg_s}{a^2}\sum\limits_{\alpha=\mathrm{c,v},\mathbf q}\left| \beta _{\alpha}(\mathbf q,t) \right|^2\mathbf v^\mathrm{(\alpha )}{(\mathbf k(\mathbf q,t))}~,
\label{intra}
\end{equation}
where $\mathbf v_{\mathbf  k}^\mathrm{(\alpha)}=\frac{\partial}{\partial\mathbf k}E^\mathrm{(\alpha)}(\mathbf k)$ is the group velocity (intraband velocity) and $g_s=2$ is the spin degeneracy. The group velocities can be found from Eqs. (\ref{Energyc})-(\ref{Energy})
\begin{eqnarray}
V_x^{\mathrm{ c}}(\mathbf k)&=&-V_x^{\mathrm{v}}(\mathbf k)=\frac{-a\gamma ^2}{\hbar{\sqrt{|\gamma{f(\mathbf k)}|^2+\frac{\Delta ^2}{4}}}}
\nonumber
\\
&\times&\sin{\frac{ak_x}{2}}\Big(\cos{\frac{\sqrt{3}ak_y}{2}}
+2\cos{\frac{ak_x}{2}}\Big)\\
V_y^{\mathrm{ c}}(\mathbf k)&=&-V_y^{\mathrm{v}}(\mathbf k)=\frac{-\sqrt{3}a\gamma ^2}{\hbar{\sqrt{|\gamma{f(\mathbf k)}|^2+\frac{\Delta ^2}{4}}}}
\nonumber
\\
&\times&\sin{\frac{\sqrt{3}ak_y}{2}}\cos{\frac{ak_x}{2} }.
\end{eqnarray}
The interband current is given by the following expression 
\begin{eqnarray}
  && \mathbf J^\text{(inter)}(t)=i\frac{eg_s}{\hbar a^2}\sum _{\substack{\mathbf q\\ \alpha,\alpha^\prime=\mathrm{v,c}\\
 \alpha\ne\alpha^\prime}}\beta _{\alpha^\prime}^\ast(\mathbf q,t)\beta _{\alpha}(\mathbf q,t)\nonumber \\&&\times\exp \{ i \phi^\mathrm{(D)}_\mathrm{\alpha^\prime\alpha}(\mathbf q,t)+ i \phi^\mathrm{(B)}_\mathrm{\alpha^\prime\alpha}(\mathbf q,t)\}\nonumber \\ 
 &&\times\left[ E_{\alpha^\prime}\left(\mathbf k(\mathbf q,t)\right)-E_\alpha \left(\mathbf k(\mathbf q,t)\right)\right] \mathbfcal A^{(\alpha\alpha^\prime)}\left(\mathbf k(\mathbf q,t)\right),
 \label{J}
\end{eqnarray}
where 
\begin{eqnarray}
&&\phi^\mathrm{(D)}_\mathrm {\alpha^\prime \alpha}(\mathbf q,t)=\phi^\mathrm{(D)}_ \mathrm {\alpha}(\mathbf q,t)-\phi^\mathrm{(D)}_ \mathrm {\alpha^\prime }(\mathbf q,t),\\
&&\phi^\mathrm{(B)}_\mathrm {\alpha^\prime \alpha}(\mathbf q,t)=\phi^\mathrm{(B)}_ \mathrm {\alpha}(\mathbf q,t)-\phi^\mathrm{(B)}_ \mathrm {\alpha^\prime }(\mathbf q,t).
\label{Eq:interband Berry phase}
\end{eqnarray}

\section{Results and discussion}


In gapped graphene, sublattices $A$ and $B$ are unequivalent, which results in broken inversion symmetry. The gapped graphene is symmetric with respect to 
the $y$-axis, but there is no symmetry with respect to the $x$-axis, see Fig.\ \ref{fig:Energy}. Thus, if the linear optical pulse is polarized along the $y$-axis, then 
the CB population distribution in the reciprocal space is symmetric with respect to the $y$-axis and the electric current is generated 
only along the $y$-axis, and not along the $x$-axis. But if the pulse is polarized along the $x$-axis, the current is expected to flow both along the $x$ and $y$ directions.
Below we consider only this case, i.e., we assume that the optical pulse is polarized along the $x$-axis.   

We consider a linearly $x$-polarized ultrafast optical pulse that is applied normally on the gapped graphene monolayer and has the following waveform
\begin{equation}
F=F_0(1-2u^2)e^{-u^2} ,
\label{field}
\end{equation}
where $F_0$ is the amplitude of the pulse, $u=t/\tau$ , and $\tau =1 $ fs. We assume that the pulse is polarized along the $x$-axis. It should be mentioned that the $x$-axis 
is not the axis of symmetry of the gapped graphene, while the $y$-axis is the axis of symmetry. 

\begin{figure}
\begin{center}\includegraphics[width=0.47\textwidth]{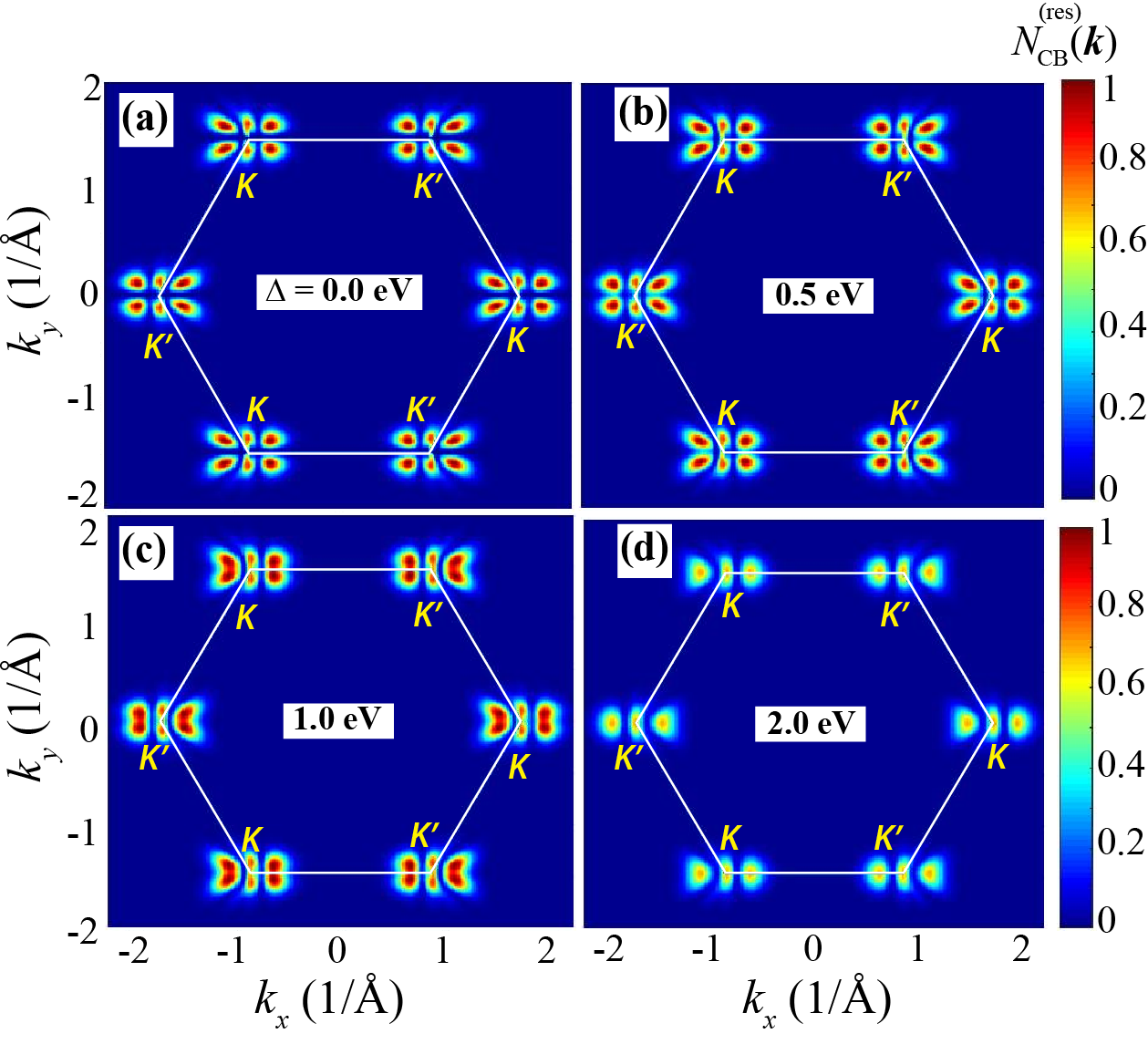}\end{center}
  \caption{(Color online) The residual CB population $N\mathrm{^{(res)}_\mathrm{CB}}(\mathbf{k})$ for gapped graphene with various bandgaps, 0 eV, 0.5 eV, 1 eV, and 2 eV, in
the extended zone picture. The white solid line shows the boundaries of the first Brillouin
zone  with  $K, K^\prime$-points indicated. The applied field is linearly polarized pulse in the x direction and its amplitude is 0.5 $\mathrm{V\AA^{-1}}$.
}
  \label{fig:NN_Linear_Fx_0p5_gap}
\end{figure}

In the presence of the pulse, we solve the TDSE assuming that the VB is initially occupied and the CB is empty. The electron dynamics in the field of the pulse is highly nonlinear and is characterized by redistribution of electrons between the valence and the conduction bands. After the pulse, there is a nonzero residual electron population, $N^\mathrm{(res)}_\mathrm{CB}$, in the CB  --see Fig.\ \ref{fig:NN_Linear_Fx_0p5_gap}. Such population 
determines the irreversibility of the electron dynamics. 
The distributions of $N^\mathrm{(res)}_\mathrm{CB}$ in the reciprocal space  are shown in Fig.~\ref{fig:NN_Linear_Fx_0p5_gap}(a)-(d)  for different values of the band gap. 
The distributions are characterized by hot spots with large, $\sim 1$, CB population.  Such hot spots are due to double passage of electrons
of the $K$ ($K^\prime$) point during the pulse and the manifestation of interference pattern. 
Similar hot spots were discussed in Ref.\onlinecite{Stockman_et_al_PRB_2016_Graphene_in_Ultrafast_Field}, where interaction of 
a linear optical pulse with pristine graphene has been studied. For gapped 
graphene, the interference pattern becomes smeared, see 
Fig.~\ref{fig:NN_Linear_Fx_0p5_gap}. This is because the interband coupling is determined by non-Abelian Berry connection, the distribution of which is 
broadened with increasing the bandgap. 
At the same time, the separation between the fringes is inversely proportional to the nonlocality distance and, thus, does not depend on the bandgap\cite{Stockman_et_al_PRB_2016_Graphene_in_Ultrafast_Field}. 
Another interesting property of the CB population distribution is that it is symmetric with respect to both $x$ and $y$ axes. This is nontrivital property since the 
the $x$-axis is not the axis of symmetry of the system.  

\begin{figure}
\begin{center}\includegraphics[width=0.47\textwidth]{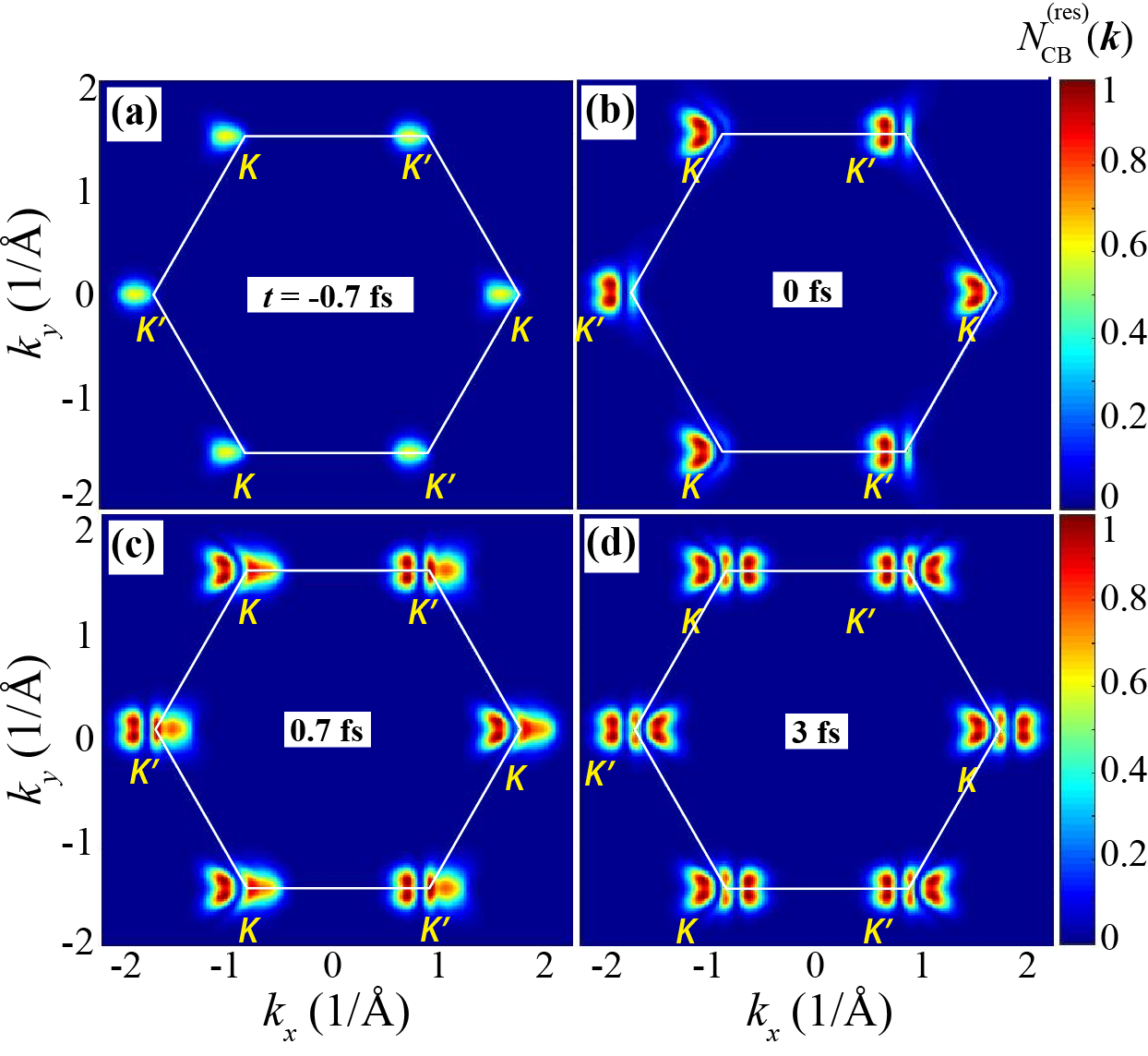}\end{center}
\caption{(Color online) The CB population $N\mathrm{_\mathrm{CB}}(\mathbf{k})$  as a function of initial lattice vector for gapped graphene with bandgap 1 eV in the extended zone picture at different moments of time, -0.7 fs, 0 fs, 0.7 fs, and 3 fs. The white solid line shows the boundaries of the first Brillouin
zone with $K, K^\prime$-points indicated. The applied pulse in linearly polarized in the x direction and its amplitude is 0.5  $\mathrm{V\AA^{-1}}$.}
\label{fig:NN_Fx_0p5_gap_1_time}
\end{figure}

The CB population distribution is shown in Fig.\ \ref{fig:NN_Fx_0p5_gap_1_time}(a)-(d) at different moments of time. It illustrates  the formation of the 
interference-induced hot spots in the CB population distribution. At all moments of time the CB population distribution is symmetric with respect to the 
$x$ axis.  
Initially, at  $-2.5~\mathrm{fs}\leq t\leq-0.7~\mathrm{fs}$, the applied field is negative so the electrons are accelerated to the right. 
Since the interband coupling is strong near the $K$ and $K^\prime$ points only, the CB population within this time interval is large on left side of the 
Dirac points, see Fig. \ref{fig:NN_Fx_0p5_gap_1_time} (a). 

For time interval $-0.7~\mathrm{fs}\leq  t\leq 0~\mathrm{fs}$, the field is positive and the electrons move to the left and pass the Dirac points the 
second time, which results in 
interference fringes or hot spots on the left sides of the valleys as shown in Fig.\ref{fig:NN_Fx_0p5_gap_1_time}(b). The field remains positive for 
 $0~\mathrm{fs}\leq  t\leq 0.7~\mathrm{fs}$ and now the electrons from the right side of the Dirac points pass the region near the $K$ or $K^\prime $ points, 
 which results in large CB population on the right side of the $K$ and $K^\prime $ points, see Fig.\ref{fig:NN_Fx_0p5_gap_1_time}(c). 
The field changes its sign at $0.7~\mathrm{fs}\leq  t\leq 2.5~\mathrm{fs}$. Then the electrons from the right side of the Dirac points pass 
through the region of large interband coupling the second time, which produce hot spots of CB population on the right side of the $K$ and $K^\prime $ points. 
The electron CB distributions shown in Figs.\ \ref{fig:NN_Linear_Fx_0p5_gap} and \ref{fig:NN_Fx_0p5_gap_1_time} could be observed by the time resolve angle-resolved photoelectron spectroscopy (tr-ARPES) \cite{Chiang_et_al_PhysRevLett.107_Berry_Phase_in_Graphene_ARPES,Freericks_et_al_Annal_Phys_2017_Superconductors_TR-ARPES_Theory}.

 \begin{figure}
\begin{center}\includegraphics[width=0.47\textwidth]{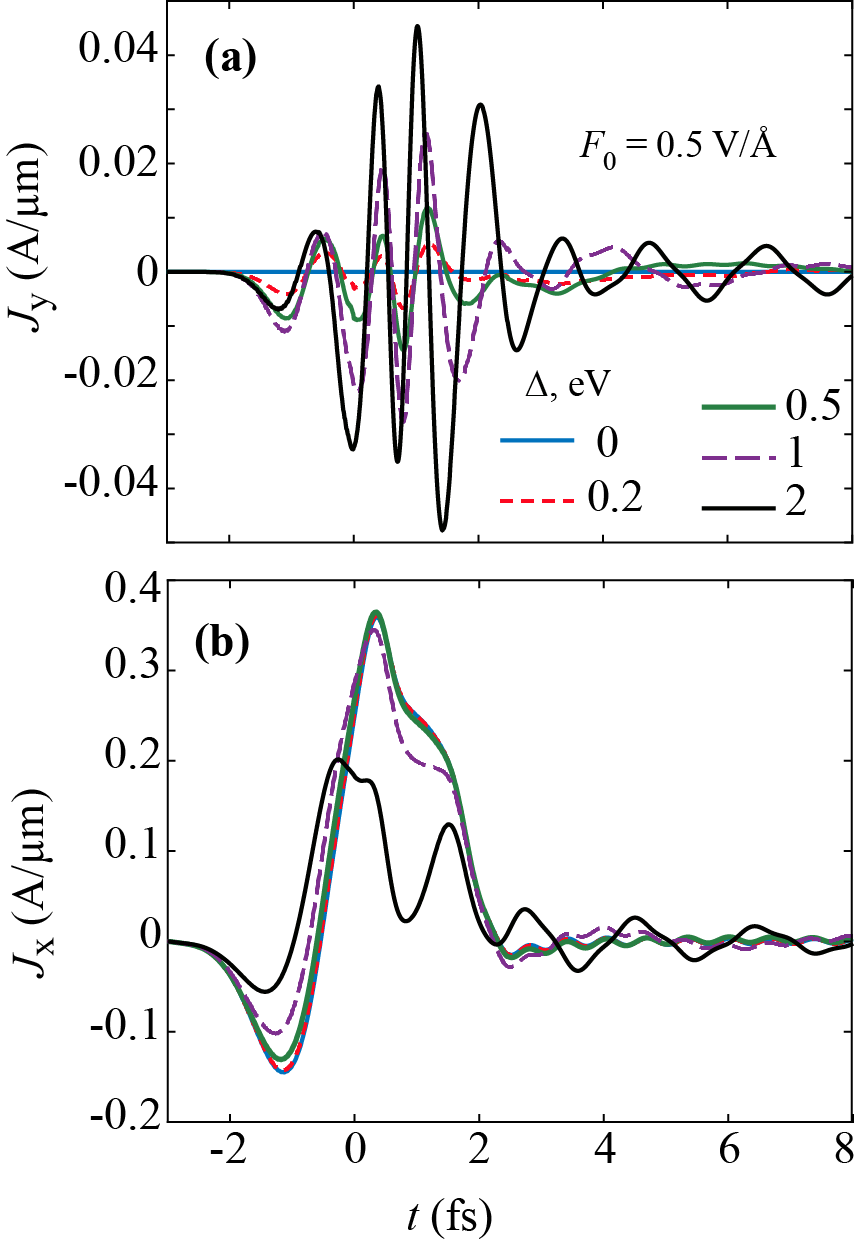}\end{center}
  \caption{The current densities in gapped graphene are shown
as a function of time for various bandgaps, 0 eV, 0.2 eV, 0.5 eV, 1 eV, and 2 eV. (a) The current density, $J_\mathrm y$, is in the direction of normal to the applied field. (b) The current density, $J_\mathrm x$, is along the direction of the applied field. The applied pulse in linearly polarized in the x direction and its amplitude is 0.5  $\mathrm{V\AA^{-1}}$.}
  \label{fig:NN_Jx_Jy_Fx_0p5}
\end{figure}%

 \begin{figure}
\begin{center}\includegraphics[width=0.47\textwidth]{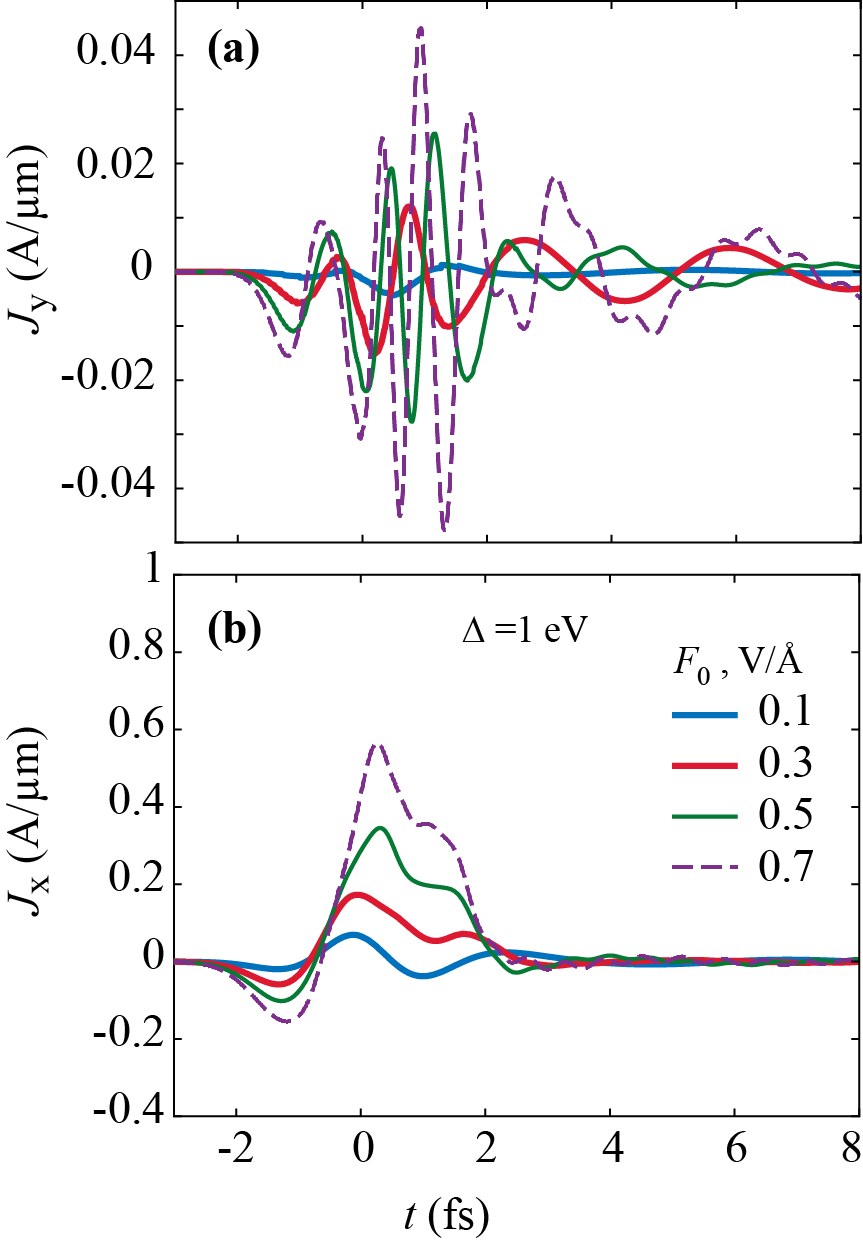}\end{center}
  \caption{The current densities in gapped graphene are shown 
as a function of time for various field's amplitudes, 0.1 $\mathrm{V\AA^{-1}}$, 0.3 $\mathrm{V\AA^{-1}}$, 0.5 $\mathrm{V\AA^{-1}}$, and 0.7 $\mathrm{V\AA^{-1}}$. (a) The current density, $J_\mathrm y$, is in the direction of normal to the applied field. (b) The current density, $J_\mathrm x$, is along the direction of the applied field. The applied pulse is linearly polarized in the x direction and the bandgap of gapped graphene is 1  $\mathrm{eV}$.}
  \label{fig:NN_Jx_Jy_Fx_gap_1eV}
\end{figure}%

Redistribution of electrons between the VB and CB during the pulse generates an electric current. For the pulse polarized along the $x$-axis, 
which is not the axis of symmetry for the gapped graphene, both the longitudinal current, i.e., the current in the $x$ direction, and the transverse 
current, i.e., the current in the $y$ direction, are generated. Such currents are shown in Fig.\ \ref{fig:NN_Jx_Jy_Fx_0p5} for different 
values of the bandgap, $\Delta$. For zero bandgap, i.e., for pristine graphene, the transverse current is zero. The transverse current increases with 
the bandgap. The electric current, generated during the pulse, has two contributions: intraband and interband. The intraband current is completely 
determined by the electron density distributions in the CB and VB. It can be also considered as a measure of asymmetry of such distributions. 
Such the CB population distribution is symmetric with respect to the $y$-axis both during the pulse and after the pulse, the intraband transverse 
current, $J_y$, is zero. Thus, the transverse current for gapped graphene is determined by the interband contribution only. As the results, the 
transverse current as a function of time is oscillating with the frequency that depends on the bandgap, see Fig.\ \ref{fig:NN_Jx_Jy_Fx_0p5}(a). 
At the same time, the longitudinal current is almost unidirectional with small oscillations, see Fig.\ \ref{fig:NN_Jx_Jy_Fx_0p5}(b). 

Since the bandgap determines the strength of the asymmetry of the system, we expect that the magnitude of the 
 transverse current increases with the bandgap, which is shown in Fig.\ \ref{fig:NN_Jx_Jy_Fx_0p5}(a). For the longitudinal current, 
there is a different tendency. The longitudinal current first increases with $\Delta $ and then at 
large bandgaps, $\Delta \sim 2 eV$, decreases. Such suppression of the longitudinal current at large values of $\Delta $ is due to the 
specific dependence of the 
interband dipole matrix elements (non-Abelian Berry connection) on the bandgap. At small bandgaps, the interband dipole matrix element is strongly localized 
near the $K$ and $K^\prime $ points. With increasing the bandgap, the dipole matrix element becomes delocalized and nonzero at large part of the 
Brillouin zone, where the maximum of the dipole matrix element decreases with the bandgap keeping the net dipole matrix element, i.e., the integral of the dipole 
matrix element over the whole Brillouin zone, constant. As a results the total CB population near the $K$ or $K^\prime $ points decreases with $\Delta $, 
which finally results in suppression of the longitudinal current. 

In Fig.\ \ref{fig:NN_Jx_Jy_Fx_gap_1eV} the longitudinal and transverse currents are shown for different field amplitudes. As expected, with increasing the 
field amplitude, the magnitudes of both currents increase. The frequency of oscillations of the transverse current also shows the 
dependence on the magnitude of the pulse, while the longitudinal current is almost unidirectional.

The direction of the current is determined by the direction of the field maximum. For the field profile (\ref{field}), the field maximum is pointing in the 
positive direction of the $x$-axis. If we change the direction of the field maximum to the negative one, i.e., it is 
pointing in the negative direction of the $x$-axis, then the longitudinal current, $J_x$, changes its sign, while the transverse current, $J_y$, 
remains the same. The transverse current changes its sign if we change the signs of the on-site energies of sublattices $A$ and $B$, i.e., change the sign 
of parameter $\Delta $ in Hamiltonian (\ref{H0}).

\begin{figure}
\begin{center}\includegraphics[width=0.47\textwidth]{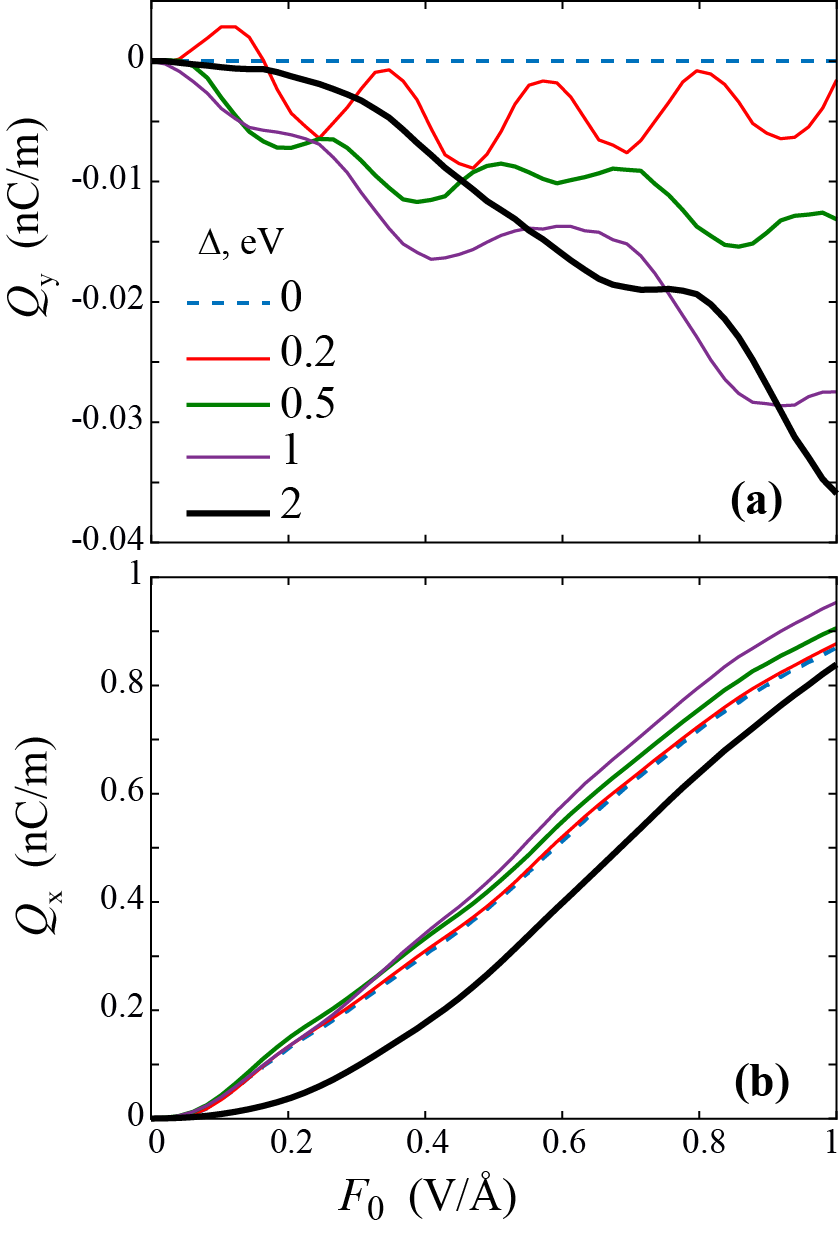}\end{center}
  \caption{The transferred charge densities are shown as a function of the field amplitude, $F_0$, for different bandgaps, 0 eV (case of graphene), 0.2 eV, 0.5 eV, 1 eV, and 2 eV. (a) The transferred charge density is shown in the direction normal to the applied field, $Q_\mathrm y$, and (b) The transferred charge density is shown along the direction of the field, $Q_\mathrm x$. The applied pulse is linearly polarized in the x direction.}
  \label{fig:NN_Qx_Qy_Fx_gap}
\end{figure}

The generated electric current during the pulse results in the transfer of an electric charge through the system. Such transferred charge can be calculated as 
\begin{equation}
\mathbf{Q}=\int_{-\infty}^\infty \mathbf{J}(t) dt.
\end{equation}
For the pulse polarized along the $x$-axis, the charge is transferred in both $x$ and $y$ directions. In Fig.\ \ref{fig:NN_Qx_Qy_Fx_gap} the transferred charge
is shown as a function of the pulse amplitude for different values of the bandgap. As expected, for zero bandgap, there is no charge transfer in the 
transverse direction, $Q_y=0$. As a function of the field amplitude, the transverse transferred charge shows oscillations, which is due to oscillations in the 
transverse current as a function of time. The longitudinal transferred charge, $Q_x$, monotonically increases with the field amplitude and has weak dependence on the 
bandgap. At large bandgap, $\Delta \sim 2 eV$, transferred charge $Q_x$ becomes smaller, which is related to suppression of the CB population and 
correspondingly the longitudinal electric current at large $\Delta $.

Changing the direction of the applied field and applying it in -x direction changes the sign of the longitudinal current  however, it does not have any effect on the normal current. This current only changes the sign if we change the on-site energies of different sublattices.


\section{conclusion}

In pristine graphene, which has an inversion symmetry, there are two axes of symmetry, say $x$ and $y$. If an external linear pulse is polarized along 
these two directions, then it will produce CB population distribution that is symmetric with respect to the axis of polarization of the pulse. The 
pulse will also generate an electric current and the corresponding transferred charge  along the direction of polarization only, but not in the transverse 
direction. 

For gapped graphene, the inversion symmetry is broken. In this case there is only one axis of symmetry, say the $y$-axis. If the linear 
pulse is polarized along the $x$ axis, then since this axis is not the axis of symmetry, the electric current is generated in both $x$ and 
$y$ directions. The transverse current does not depends on the direction of the field maximum, while the longitudinal current changes 
its sign when the direction of the maximum is reversed. At the same time, for the same polarization of the pulse, i.e., along the $x$-axis,
similar to pristine graphene, the CB population distribution is 
symmetric with respect to the $x$-axis both during the pulse and after the pulse. 
It means that the electron dynamics above ($k_y>0$) and below ($k_y<0$) 
the $K$ ($K^\prime$) point is exactly the same, which results in symmetric CB population distribution. Although the electron dynamics depends on 
the geometric phase, which is different above and below the $K$ ($K^\prime $) point, this phase is exactly canceled by the phase of the 
interband dipole matrix element (non-Abelian Berry connection). This is the property of the two-band model of gapped graphene which will be discussed somewhere else. If more bands are 
included into the model, then there will be no cancellation of the geometric phase and the net (topological) phase, which is the sum of the 
geometric phase and the phase of the interband dipole coupling, will be nonzero. The topological phase has different time dependence above and 
below the $K$ ($K^\prime $), which results in topological resonance. The topological resonance occurs due to a partial cancellation of the dynamic 
phase by the topological phase. Such partial cancellation is different above and below the $K$ ($K^\prime$) point, which finally results in 
different CB populations and asymmetric CB population distribution.  
Such small asymmetry of CB population will introduce small intraband contribution to the transverse current.

\begin{acknowledgments}
Major funding was provided by Grant No. DE-FG02-11ER46789 from the Materials Sciences and Engineering Division of the Office of the Basic Energy Sciences, Office of Science, U.S. Department of Energy. Numerical simulations have been performed using support by
Grant No. DE-FG02-01ER15213 from the Chemical Sciences, Biosciences and Geosciences Division, Office of Basic Energy Sciences, Office of Science, US Department of Energy. The work of V.A. was supported by NSF EFRI NewLAW Grant EFMA-17 41691. Support for S.A.O.M. came from a MURI Grant No. FA9550-15-1-0037 from the US Air Force of Scientific Research.
\end{acknowledgments}
 

%

\end{document}